\documentclass{article}

\usepackage{titling}

\usepackage{multicol}
\usepackage{wrapfig}
\usepackage{authblk}
\usepackage{float}
\usepackage{rotating}
\usepackage{lscape}
\usepackage{pdflscape}
\usepackage[skins]{tcolorbox}
\usepackage{graphicx}
\usepackage{xspace}
\usepackage{sectsty}
\usepackage{soul}
\usepackage{url}
\makeatletter
\g@addto@macro{\UrlBreaks}{\UrlOrds}
\makeatother

\title{Cloud Futurology}

\usepackage[sc]{mathpazo} 
\usepackage[T1]{fontenc} 
\usepackage{microtype} 
\usepackage[hmarginratio=1:1,top=25mm,columnsep=20pt]{geometry} 
\usepackage[hang, small,labelfont=bf,up,textfont=it,up]{caption} 
\usepackage{booktabs} 
\usepackage{float} 
\usepackage{paralist} 

\usepackage{abstract} 

\usepackage{titlesec} 
\renewcommand\thesubsection{\Roman{subsection}} 
\titleformat{\section}[block]{\large\scshape\centering}{\thesection.}{1em}{} 
\titleformat{\subsection}[block]{\large}{\thesubsection.}{1em}{} 

\usepackage{fancyhdr} 
\usepackage{amsfonts,amsmath,amssymb,amstext,latexsym}
\usepackage{color}
\usepackage{xcolor}

\usepackage[pdftex,
            pdfauthor={Yehia Elkhatib},
            pdftitle={Cloud Futurology},
            pdfkeywords={Cloud computing, Fog computing, Micro-clouds, Mini-clouds, Cloudlets, Internet of Things}]{hyperref}

\usepackage{xspace}

\makeatletter
\newcommand*{\etc}{%
    \@ifnextchar{.}%
        {etc}%
        {etc.\@\xspace}%
}
\newcommand*{\etal}{%
    \@ifnextchar{.}%
        {et al}%
        {et al.\@\xspace}%
}
\makeatother

\begin{document}

\date{}
\title{\thetitle}

\author{\textbf{\textcolor{red}{Pre-print version; accepted to IEEE Computer, 2019}}
\\
Blesson Varghese\footnote{Corresponding Author; E-mail: b.varghese@qub.ac.uk}, Philipp Leitner, Suprio Ray, Kyle Chard, Adam Barker, \\Yehia Elkhatib, Herry Herry, Cheol-Ho Hong, Jeremy Singer, Fung Po Tso,\\Eiko Yoneki, Mohamed-Faten Zhani
}

\maketitle

\thispagestyle{fancy} 

\begin{abstract}
The Cloud has become integral to most Internet-based applications and user gadgets. This article provides a brief history of the Cloud and presents a researcher's view of the prospects for innovating at the infrastructure, middleware, and application and delivery levels of the already crowded Cloud computing stack. 
\end{abstract}

\section*{Introduction}
The global Cloud computing market exceeds \$100 billion and research in this area has rapidly matured over the last decade. During this time many buzz words have come and gone. 
Consequently, traditional concepts and conventional definitions related to the Cloud are almost obsolete~\cite{nextgenclouds-1}. 
The current Cloud landscape looks very different from what may have been envisioned during its inception. 
It may seem that the area is saturated and there is little innovative research and development to be done in the Cloud, which naturally raises the question - \textit{`What is the future of the Cloud?'}

This article first examines the multiple generations of innovation that the Cloud has undergone in the last decade, it then presents a researcher's view of the prospects and opportunities for innovation in this area across the entire Cloud stack - highlighting the infrastructure, middleware, and application and delivery levels.

\section*{The Cloud Landscape}
\label{sec:landscape}

\begin{sidewaysfigure*}
	\centering
	\includegraphics[width=\textwidth,keepaspectratio]{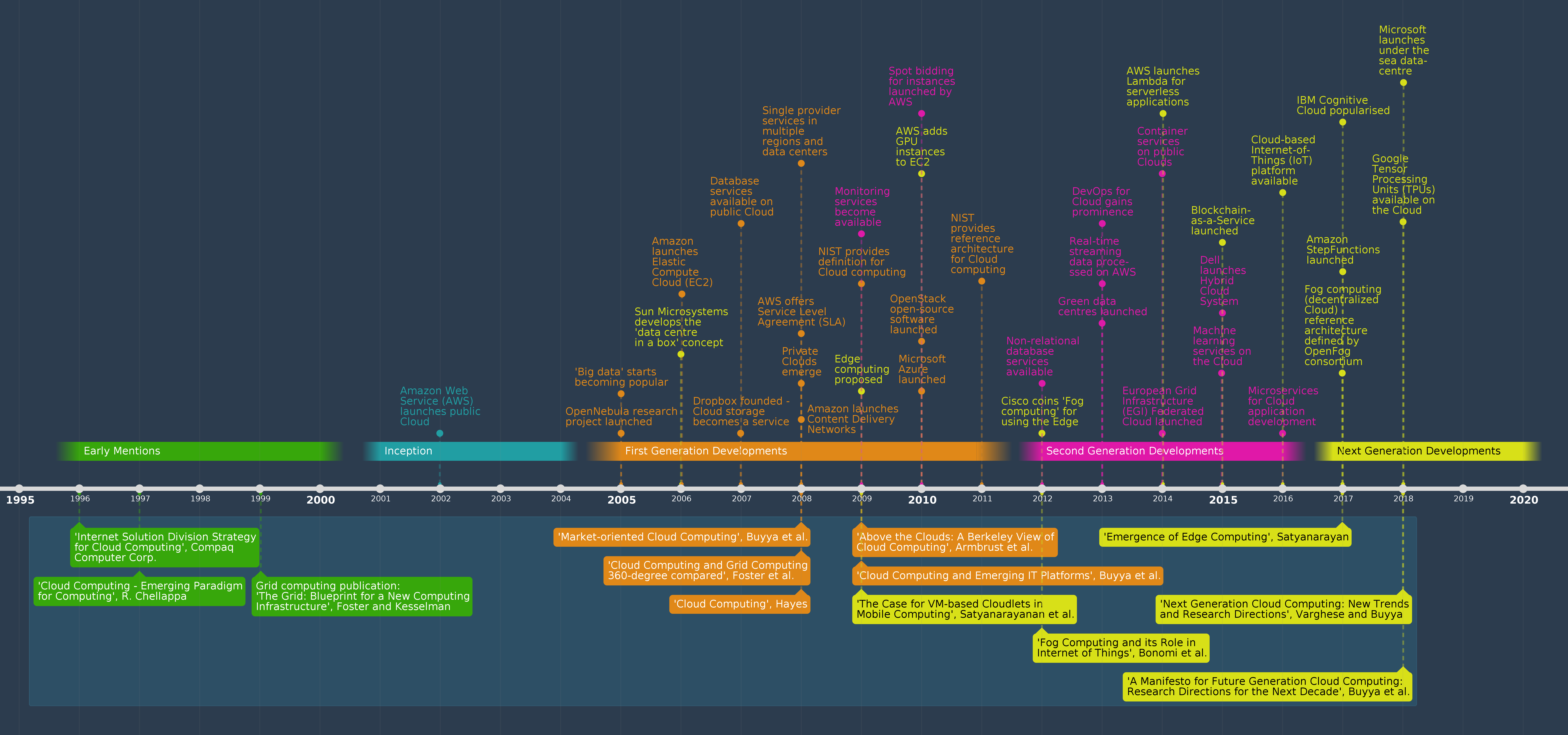}
	\caption{A timeline of the Cloud computing landscape. {Early mentions of the Cloud in literature are described in the green block. The period of inception is highlighted in turquoise. This is followed by two generations of developments on the Cloud - the orange block highlights first generation and the purple colour presents second generation developments. Upcoming developments and trends are presented as next generation developments in the yellow block.}}
	\label{fig:timeline}
\end{sidewaysfigure*}

\begin{tcolorbox}[
	enhanced,
	drop fuzzy shadow southwest,
	colframe=red!50!black,
	colback=yellow!10,]
	The first known suggestion of Cloud-like computing was by Professor John McCarthy at MIT's centennial celebration in 1961 \textit{``Computing may someday be organized as a public utility just as the telephone system is a public utility ... Each subscriber needs to pay only for the capacity he actually uses, but he has access to all programming languages characteristic of a very large system ... Certain subscribers might offer service to other subscribers ... The computer utility could become the basis of a new and important industry''}
\end{tcolorbox} 

{Two fundamental technologies required for developing the Cloud, namely virtualization and networking, were first developed in the 60's. In 1967, IBM virtualized operating systems to allow multiple users to share the same computer and in 1969, the US Department of Defense launched the Advanced Research Projects Agency Network (ARPANET), which defined network protocols that led to the development of the Internet.}
Although the earliest mentions of Cloud computing in literature appear in the 90’s as shown in Figure~\ref{fig:timeline}, it was Grid computing~\cite{foster99grid} that laid the foundation for offering computing resources as a service to users in the 90's and early 21st century. 
The inception of the Cloud as a utility service was realized when Amazon launched its commercial public Cloud in 2002. 

Significant advances over the last decade can be divided into two generations as seen in Figure~\ref{fig:timeline}. The first generation focuses on development at the infrastructure level, for example creating data centres, which are centralized infrastructure that host significant processing and storage resources across different geographic regions. Across other layers of the Cloud stack a range of user-facing services emerged, some of which were available only in specific geographic regions. Software developed by OpenNebula (\url{http://www.opennebula.org}) and OpenStack (\url{https://www.openstack.org}) allowed organizations to own private Clouds and set up their own data centres. 

As big data started to gain popularity in 2005, the Cloud was a natural first choice to tackle big data challenges. This led to the popularity of storage services such as Dropbox that relied on the Cloud. Relational databases hosted in the Cloud to support enterprise applications emerged.

Since inception in early 2000 and despite significant research and development efforts spanning over half a decade, a reference architecture for the Cloud was not defined until 2011 (\url{https://ws680.nist.gov/publication/get_pdf.cfm?pub_id=909505}). Since the Cloud was a new technology it may have taken a few years before definitions were articulated, circulated, and widely accepted. 

Second generation developments focused on enriching the variety of services and their quality. In particular, management services and modular applications emerged. 
Monitoring services of compute, network and storage resources offering aggregate and fine metrics became available to application owners, allowing them to maximize performance.
More flexible pricing strategies and 
	service level agreements (SLAs), in addition to the posted price, pay-as-you-go model, such as spot bidding and preemptible virtual machine instances emerged in 2010. 

Furthermore, the move from immutable virtual machines to smaller, loosely coupled execution units in the form of microservices and containers was a game changer for decomposing applications within and across different data centres. Combining public and private (on-premise) Clouds of different scale (a.k.a \textit{cross-cloud or {hybrid Cloud computing}}~\cite{Elkhatib2016crosscloudmap}) gained prominence in order to alleviate concerns related to privacy and vendor lock-in.

An important step in the evolution of the Cloud was the development of Content Delivery Networks (CDNs). Compute and storage resources were geographically distributed for improving the overall quality of a variety of services, including streaming and caching. CDNs are the basis of upcoming trends in decentralizing Cloud resources towards the edge of the network, which will be considered in this article. Amazon launched their CDN in 2008.

\begin{tcolorbox}[
	enhanced,
	drop fuzzy shadow southwest,
	colframe=red!50!black,
	colback=yellow!10,]
	Containers are namespaces with ring-fenced resources. For example, Docker~\cite{merkel2014docker} is a popular container technology for creating and managing self-sufficient execution units. Containers offer the prospect of seamless application development, testing, and delivery over heterogeneous environments.
	The microservice architectural style focuses on how the application logic is implemented rather than how it is hosted by dividing services into atomic functions in order to tame operational complexity. The emphasis is on developing small, replaceable service units instead of maintaining monolith services~\cite{nadareishvili2016microservice}.
\end{tcolorbox}

\section*{Innovation for the Next Generation}
\label{sec:innovation}
Tangential innovations in the first and second generation developments have made way for another generation of Cloud development that will focus on decentralization of resources.
Although there has been a decade long explosion of Cloud research and development, there is significant innovation yet to come in the infrastructure, middleware, and application and delivery areas.  

As shown in the `Next Generation Development' block of Figure~\ref{fig:timeline} in the next five years, computing as a utility will be miniaturized and available outside large data centres. Referred to as 'Cloud-in-a-Box', Sun Microsystems first demonstrated these ideas in 2006 and paved the way for Fog/Edge computing. The use of hardware accelerators, such as Graphics Processing Units (GPUs), in the Cloud began in 2010 is now leading to inclusion of even more specialized accelerators that are, for example, customized for modern machine learning or artificial intelligence workloads. Google provides Tensor Processing Units (TPUs) that are customized for such workloads with the aim to deliver new hardware and software stacks that extend machine learning and artificial intelligence capabilities both within and outside the cloud.

\subsection*{Infrastructure}
\label{sec:infrastructure}

A range of hardware accelerators, such as Graphics Processing Units (GPUs), Field-Programmable Gate Arrays (FPGAs), and more specialized Tensor Processing Units (TPUs) are now available for improving the performance of applications on the Cloud. Typically, these accelerators are not shared between applications and therefore result in an expensive data centre setup given the large amount of underutilized hardware.

\textit{\textbf{Accelerator virtualization}} is the underlying technology that allows multiple applications to share the same hardware accelerator~\cite{hong2017gpu}. All existing virtualization solutions have performance limitations and are bespoke to each type of accelerator. Given that this is a relatively new area of study, we are yet to see a robust production-level solution that can easily incorporate and virtualize different types of accelerators with minimal overheads.

Currently, applications leverage Cloud resources from geographically distant data centres. Consequently, a user device, such as a smartphone must transfer data to a remote Cloud for processing data. However, this will simply not be possible when billions of devices are connected to the Internet, as frequent communication and communication latencies will affect the overall service quality and experience of the user. These latencies can be reduced by bringing compute resources to the network edge in a model often referred to as \textbf{\textit{Fog/Edge computing}}~\cite{edgecomputing-02, edgecomputing-01}.

Edge computing is challenging - Edge resources need to be publicly and securely available. However, the risks, security concerns, vulnerabilities and pricing models are not articulated, or even fully known. Implementations of standardized Edge architectures and a unified marketplace is likely to emerge. 
Furthermore, there are no robust solutions to deploy an application across the Cloud and the Edge. Toolkits for deploying and managing applications on the Edge will materialize given the community led efforts by the European Telecommunications Standards Institute, the OpenFog consortium, and OpenStack.

\vspace{2pt}
\begin{tcolorbox}[
	enhanced,
	drop fuzzy shadow southwest,
	colframe=red!50!black,
	colback=yellow!10,]
	Fog/Edge computing~\cite{edgecomputing-02, edgecomputing-01} refers to the use of resources at the Cloud to the network edge continuum. Data from user devices can be processed at the edge instead of remote data centers. Gartner 
	and Forester 
	define Edge computing as an enabling and strategic technology for realizing IoT and for building future Cloud applications. The Edge computing market is estimated to be worth US \$6.72 billion within the next five years with a compound annual growth rate of over 35\%.
\end{tcolorbox}

\begin{figure}[H]
	\centering
	\includegraphics[width=0.75\textwidth]{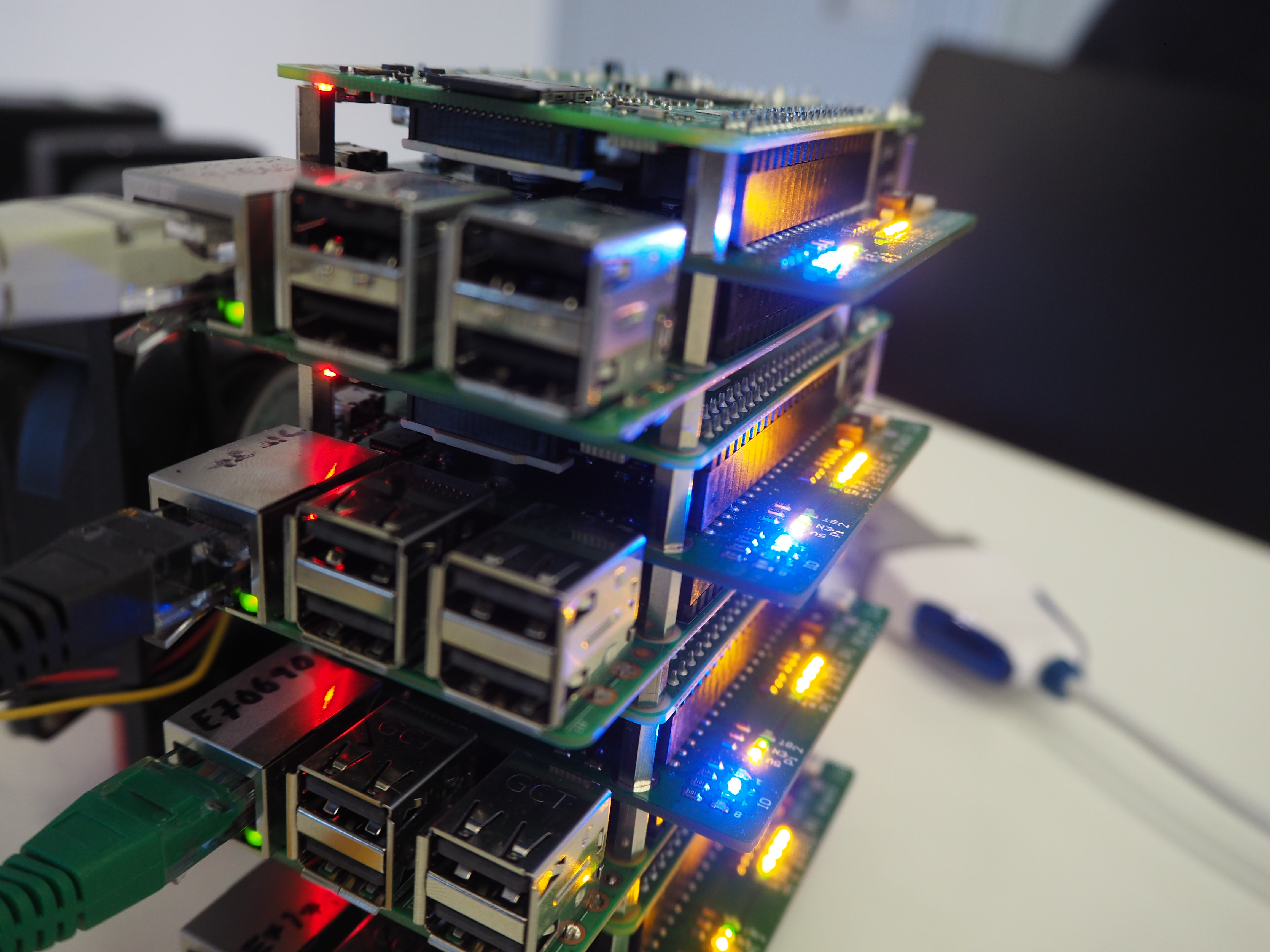}
	\caption{A micro data centre comprising a cluster of Raspberry Pis that was assembled at the University of Glasgow, UK. These data centres in contrast to large Cloud data centres are low cost and low power consuming.}
	\label{fig:figure1}
\end{figure}

The definition of Cloud data centres is also changing - what was conventionally `dozens of data centres with millions of cores,' is now moving towards `millions of data centres with dozens of cores.' These \textbf{\textit{micro data centres}} are novelty architectures and miniatures of commercial cloud infrastructure {(see Figure~\ref{fig:figure1})}. Such deployments have become possible by networking several single board computers, such as the popular Raspberry Pi~\cite{Cox2013}. Consequently, the overall capital cost, physical footprint and power consumption is only a small fraction when compared to the commercial macro data centres~\cite{elkhatib2017microclouds}. 

\vspace{2pt}
\begin{tcolorbox}[
	enhanced,
	drop fuzzy shadow southwest,
	colframe=red!50!black,
	colback=yellow!10,]
	Micro data centres are a compelling infrastructure to support education, where students can be exposed to near-real data centre scenarios in a sandbox. They are widely used in educational contexts and more convincing use-cases will emerge. Global community engagement will facilitate the further adoption of micro data centres. International competitions and league tables for micro data centre deployment would catalyze this process.
\end{tcolorbox} 

The challenges to be addressed in making micro data centres operational include reducing overheads of the software infrastructure. For example, machine learning libraries must be optimized for use on micro data centre hardware. Additionally, management tools must run on devices that have only limited network access, perhaps due to the presence of firewalls, or intermittent connectivity as commonly found on edge networks. This is different from traditional data centres, which operate on the assumption that managed nodes are directly and permanently reachable.

{ Data centers are now one of the largest consumers of energy (\url{http://www.climatechangenews.com/2017/12/11/tsunami-data-consume-one-fifth-global-electricity-2025/}). This is in part due to the end of Moore's law and the fact that increasing processor speed no longer offsets energy consumption. It is also due to the rapid growth of Internet-of-Things (IoT) - with estimates of up to 80 billion devices online by 2025 - and increasing use in developing countries. 
	As such, Cloud providers are facing both economic and legislative pressures to reduce energy consumption. 
	It is unlikely that new power plants will be sufficient to meet these growing energy needs and thus there will be widespread use of renewable energy sources and `stranded power'~\cite{yang16zccloud}. 
}

\subsection*{Middleware}
\label{sec:middleware}

Cloud customers are spoiled for choice, to the extent that it has become overwhelming for many. This is because of the incredible rate at which the Cloud resource and service portfolio has expanded. As such, there is now a need not only for middleware technologies that abstract differences between Clouds and services, but also for decision support systems to aid customer deployment of applications. For example, to help guide selection of the best Cloud providers, resources and services, and configuration. 

Cross-cloud challenges, such as identifying optimal deployment resources from across the vast array of options from different providers, seamlessly moving workloads between providers, and building systems that work equally well across the services of different providers will need to be surmounted for establishing a viable \textbf{\textit{Cloud federation}}.

Such systems, naturally, cannot be one-size-fits-all, but they must be tailored to the needs of the customer and to follow any changes in the Cloud provisioning market. 
\textbf{\textit{Resource brokers}} are likely to become necessary to fill this void and provide a means of exploiting differences between Cloud providers, and identifying the real performance profiles of different Cloud services before matching them to the customer needs. Currently however, no practical brokering solutions are available.

Brokers can also be used to automatically configure the parameters of applications for a set of selected resources so as to maximize the performance of applications. Typically, the configuration parameters are manually tuned, which is cumbersome given the plethora of resources. A generic \textbf{\textit{auto-tuner}} that operates in near real-time and measures performance cost-effectively is ideal. 
However, measuring performance in the Cloud is time consuming and therefore expensive~\cite{cloudbench-1,cloudbench-2}. 
In a transient environment, such as the Cloud, the performance metrics will be obsolete if they cannot be gathered in real time. The complexity increases when resources from multiple providers are considered. A starting point may be to develop bespoke auto-tuners based on domain specific knowledge that a system engineer may possess using a combination of machine learning techniques, such as deep neural networks and reinforcement learning~\cite{BOAT}.

While resource brokerage allows customers to select and combine services to their liking, Cloud data centre operators can also choose a variety of Network Functions (NFs) to create in-network services and safeguard networks to improve an application's performance. The process of orderly combining different combination or subsets of NFs, such as firewalls, network monitors, and load balancers, is referred to as \textbf{\textit{service chaining}}~\cite{cui2018enforcing}. 

Service chaining will be particularly useful for Edge computing to{, for instance,} improve data privacy. 
{Recent proposals of intent-driven networking~\cite{elhabbash2018mediation} allow operators and end-users to define \textit{what} they want from the network, not \textit{how}. This enables the on demand composition of bespoke network logic, allowing much more refined application control and dynamicity.}
A research challenge here is to manage services across Cloud and Edge networks and resources that have different ownership and operating objectives. 

\vspace{2pt}
\begin{tcolorbox}[
	enhanced,
	drop fuzzy shadow southwest,
	colframe=red!50!black,
	colback=yellow!10,]
	An interesting starting point for implementing service chaining will be creating personalized network services across Cloud and Edge environments. This will provide, for example, a user with a personalized security profile that is adaptive to different environments, such as home or office. Network functions will need to be miniaturized for Edge resources to facilitate chaining.
\end{tcolorbox}

\subsection*{Application and Delivery}
\label{sec:application}
The Cloud will continue to be an attractive proposition for \textbf{\textit{big data applications}} to meet the volume, velocity, and variety challenges~\cite{bigdata}. Apache Spark, Hadoop MapReduce and its variants have been extensively used to process volumes of data in the last five years. However, for many users these frameworks remain inaccessible due to their steep learning curve and lack of interactivity.

Further, while frameworks such as Storm and Kafka address some of the challenges associated with data velocity {(e.g., as seen in real-time streaming required by social media, IoT and high frequency trading applications)}, scaling resources on the Cloud to meet strict response time requirements is still an active research area. {This requires complex stream processing with low latency, scalability with self-load balancing capabilities and high availability.
	Many applications that stream data may have intermittent connectivity to Cloud back-end services for data processing and it would be impossible to process all data at the edge of the network. Thus, efficient stream processing frameworks that can replicate stream operators over multiple nodes and dynamically route stream data to increase potential paths for data exchange and communication will be desirable.}

Innovation will be seen in taming traditional data challenges. For example, emergence of tools that alleviate the burden on the user, that support elasticity - dynamic provisioning and fair load balancing, and that cleverly move data. As data are increasingly large and distributed, the cost of moving data can now exceed the cost of processing it. Thus, there will be increasing interest in moving {computation} to data and constructing federated registries to manage data across the Cloud. 

{A new class of distributed data management systems, called NewSQL~\cite{Pavlo2016NewSQL}, are emerging. These systems aim to offer similar scalable performance to NoSQL while supporting Atomicity, Consistency, Isolation, and Durability (ACID) properties for transactional workloads that are typical of traditional relational databases.} However, providing ACID guarantees across database instances that are distributed across multiple data centers is challenging. This is simply because it is not easy to keep data replicas consistent in multiple data locations. More mature approaches that will allow for data consistency will emerge to support future modularization of applications on the Cloud.

{A variety of existing and upcoming applications, such as smart homes, autonomous trading algorithms and self-driving cars, are expected to generate and process vast amounts of time-series data. These applications make decisions based on inputs that change rapidly over time. Conventional database solutions are not designed for dealing with scale and easy-use of time-series data. Therefore, time-series databases are expected to become more common.}

\vspace{2pt}
\begin{tcolorbox}[
	enhanced,
	drop fuzzy shadow southwest,
	colframe=red!50!black,
	colback=yellow!10,]
	Gartner predicts that by 2022 nearly 50\% of enterprise-generated data will be processed outside the traditional Cloud data centre. Processing data outside the Cloud is the premise of Edge computing. It is likely that novel methods and tools that perform complex event processing across the Cloud and Edge will emerge.
\end{tcolorbox}

A new class of applications are starting to emerge on the Cloud, namely \textbf{\textit{serverless applications}}. They are exemplified by the Function-as-a-Service (FaaS) Cloud, such as AWS Lambda or Azure Cloud Functions.
{ FaaS Clouds are implemented on top of upcoming containerization technology, but provide convenient developer abstractions on top of, for instance, Docker.}
Contrary to traditional Cloud applications that are billed for the complete hour or the minute, serverless applications are billed by the millisecond~\cite{serverless}.
FaaS has not yet seen widespread adoption for business-critical Cloud applications. This is because FaaS services and tooling are still immature and sometimes unreliable
{ Furthermore, since FaaS Clouds rely on containers significant overheads are incurred for on-demand boot up. This may be problematic for an end-user facing use cases.} 
FaaS Clouds have limited support for reuse, abstraction and modularization, which are usually taken for granted when using distributed programming models. 
{ Another practical challenge is that current-day FaaS services are entirely stateless. All application state needs to be handled in external storage services, such as Redis. FaaS providers are currently working towards stateful storage solutions, but it is as of yet unclear what these solutions will look like in practice.}
Consequently, significant developer effort is currently required to take advantage of FaaS services.

\vspace{2pt}
\begin{tcolorbox}[
	enhanced,
	drop fuzzy shadow southwest,
	colframe=red!50!black,
	colback=yellow!10,]
	FaaS Clouds have obvious advantages and we will witness innovation at the virtualization front either to reduce the overheads of containers or in the development of entirely new light-weight virtualization technologies. More powerful programming abstractions for composing and reusing Cloud functions will emerge. 
\end{tcolorbox}

{ Delivery of cloud services via economic models is transforming computing into a commodity, inline with McCarthy's  vision of computing utility. These economic models, and benefits from economies of scale, have underpinned much of the success of the Cloud.} The Cloud now utilizes a range of economic models, from posted price models, through to dynamic, spot markets for delivering infrastructure resources and a suite of higher-level platform and software services~\cite{economics}. AWS even allows users to directly exchange reserved resources with one another via a reseller market. 
As the Cloud moves towards further decoupled resources (e.g., as seen in serverless computing) new economic models are needed to address granular resource bundles, short-duration leases, and flexible markets for balancing supply and demand. 
{ Furthermore, granular, differentiated service levels are likely to become common, enabling greater user flexibility with respect to price and service quality.
	This increasingly diverse range of economic models will further enable flexibility; however, it will require greater expertise to understand trade-offs and effectively participate in the market.} Cloud automation tools will emerge to alleviate this burden by enabling users to directly quantify and manage inherent trade-offs such as cost, execution time, and solution accuracy.

\vspace{2pt}
\begin{tcolorbox}[
	enhanced,
	drop fuzzy shadow southwest,
	colframe=red!50!black,
	colback=yellow!10,]
	Current Cloud providers operate as independent silos, with little to no ability for users to move resources between providers. From a technology standpoint, creating federated Clouds via standardized abstractions is a solution, but general markets in which Cloud offerings can be compared and delivered { are expected to appear soon}. Users may be potentially offered many interchangeable alternatives, and therefore new economic models will need to cater for competition between providers rather than consumers.
\end{tcolorbox}

\section*{Opportunities and Outlook}
\label{sec:conclusions}

This article highlights a researcher's view of the prospects at the infrastructure, middleware, and application and delivery level of the Cloud computing landscape.

{ 
	Although there are ongoing efforts to tackle research challenges at the infrastructure level in four avenues, namely accelerator virtualization, Fog/Edge computing, micro data centres, and power and energy-aware solutions, the timeline to mass adoption will vary. 
	rCUDA (http://rcuda.net/) and gVirtus (https://github.com/RapidProjectH2020/GVirtuS) are exemplars of accelerator virtualization solutions that have undergone a decade of innovation, but are yet to become available on production level Cloud systems. With the increasing number of accelerators used in the Cloud it is foreseeable that accelerator virtualization technologies are adopted within the next decade.  
	Similarly, there are limited Fog/Edge computing test-beds and real deployments of such systems are yet to be seen. While the mechanisms required to adopt Fog/Edge systems are currently unknown, the growing emphasis on 5G communication will accelerate interest in the area in the short term. 
	Most micro data centres are still prototypes\footnote{\url{https://news.microsoft.com/features/under-the-sea-microsoft-tests-a-datacenter-thats-quick-to-deploy-could-provide-internet-connectivity-for-years/}} and will become widespread as more compelling use-cases, for example in Fog/Edge computing emerge over the next five years. }

{
	At the middleware level Cloud federation, resource brokers and auto-tuners, and service chaining were considered. Cloud federation in its strictest sense, i.e. reaching operational arrangement between independent providers, has for long been discussed but never in fact emerged on the horizon. Nevertheless, efforts into tools to bridge services of different providers are expected to continue. Current resource brokers and auto-tuners focus on the challenge of abstraction, but very few tackle this along with delegation, i.e., fully adaptive life cycle management. With increasing use of machine learning, the mechanisms of enabling such smart brokerage and auto-tuning are becoming available. The related challenges of expressing and interpreting customer requirements is also becoming a significant trend towards a vision where the customer needs to know far less about the infrastructure than current DevOps. In a similar vein, capturing and satisfying what an application needs from the network as seen in service chaining is a prominent research challenge. However, network operators are not as advanced in the pay-as-you-go model as cloud providers are. We expect this to change over the next five years, opening up a wide range of resources for supporting applications across an end-to-end network.
}

{
	The four key areas identified for innovation at the applications and delivery level are complex stream processing, big data applications and databases, serverless applications and economic models. 
	The rapid increase in connected devices, IoT, and streaming data is driving the immediate development of new real-time stream processing capabilities. In the next five years, we expect to see a 
	variety of new stream processing techniques, including those designed to leverage edge devices. 
	Big data applications and databases have been the focus of recent research and innovation, and thus much immediate focus is on the adoption and use of these efforts. In the next 2-5 years, there 
	will be increasing effort focused on time-series databases (following on from the release of Amazon Timestream and Azure Time Series Insights). 
	Serverless  computing is still in its infancy and is likely to be the most disruptive of the changes at the applications and delivery level. Over the next ten years, serverless technologies will likely
	permeate IT infrastructure, driving enhancements in terms of performance, flexibility (for example, support for more general application types), and economic models. 
	Economic models will evolve over the next 2-5 years due to maturing serverless and decoupled computing infrastructure. The underlying models will account for energy policies and the desire for more flexible SLAs, and will leverage new opportunities for intercloud federations. 
}

We recommend research focus in designing and developing the following important areas:

\begin{itemize}
	\item Novel lightweight virtualization technologies for workload specific accelerators, such as FPGAs and TPUs, that will proliferate the Cloud to Edge continuum and facilitate low overhead serverless computing. 
	
	\item System software of micro data centres for remote management of the system stack and workload orchestration. 
	
	\item {New power-aware strategies at the middleware and application levels for reducing energy consumption of data centres.}
	
	\item A unified marketplace for Fog/Edge computing that will cater for competition between providers rather than consumers, and techniques for adopting Fog/Edge computing rather than simply making devices and resource Fog/Edge enabled.
	
	\item Common standards for Cloud federation to allow users to freely move between providers and avoid vendor lock-in. This includes, developing interoperable systems that enforce user policies, manage the lifecycle of workloads, and negotiate Service Level Agreements (SLAs). 
	
	\item Mechanisms for vertically chaining network functions across different Cloud and Edge networks, multiple operators and heterogeneous hardware resources.  
	
	\item Novel techniques for spatio-temporal compression to reduce data sizes, adaptive indexing to managing indexes while processing queries, and developing `NewSQL' systems to support both transactional and analytical workloads on the Cloud.
\end{itemize}

Although Cloud computing research has matured over the last decade, there are numerous opportunities to pursue meaningful and impactful research in this area.

\section*{Acknowledgements}
Thanks to Dr Colin Perkins from the University of Glasgow for providing the photo in Figure~\ref{fig:figure1}.

\bibliographystyle{IEEEtran}
\bibliography{references}

\end{document}